\documentclass[aps,twocolumn,prb,showpacs,floatfix,amsmath,amssymb,superscriptaddress]{revtex4-1}
\usepackage{graphicx}
\usepackage{bm}
\usepackage{hyperref}
\usepackage{color}
\usepackage{natbib}
\usepackage{changes,cancel,amsmath}
\usepackage{braket}

\DeclareMathOperator{\tr}{Tr}

\begin{document}
\title{Effective metal-insulator non-equilibrium quantum phase transition}

\author{S. Porta}
\affiliation{Dipartimento di Fisica, Universit\`a di Genova, 16146 Genova, Italy}
\affiliation{SPIN-CNR, 16146 Genova, Italy}

\author{N. Traverso Ziani}
\affiliation{Institute for Theoretical Physics and Astrophysics, University of W\"urzburg, 97074 W\"urzburg, Germany}
\affiliation{Dipartimento di Fisica, Universit\`a di Genova, 16146 Genova, Italy}

\author{D. M. Kennes}
\affiliation{Dahlem Center for Complex Quantum Systems and Fachbereich Physik, Freie Universit\"at Berlin, 14195 Berlin, Germany}

\author{F. M. Gambetta}
\affiliation{School of Physics and Astronomy, University of Nottingham, Nottingham, NG7 2RD, UK}
\affiliation{Centre for the Mathematics and Theoretical Physics of Quantum Non-equilibrium Systems, University of Nottingham, Nottingham, NG7 2RD, UK}

\author{M. Sassetti}
\affiliation{Dipartimento di Fisica, Universit\`a di Genova, 16146 Genova, Italy}
\affiliation{SPIN-CNR, 16146 Genova, Italy}

\author{F. Cavaliere}
\affiliation{Dipartimento di Fisica, Universit\`a di Genova, 16146 Genova, Italy}
\affiliation{SPIN-CNR, 16146 Genova, Italy}

\begin{abstract}
We consider the steady state behavior of observables in the Su-Schrieffer-Heeger model and in the one dimensional transverse field quantum Ising  model after a sudden quantum quench of the parameter controlling the gap. In the thermodynamic limit, and for times $t\rightarrow\infty$, we find non-analyticities even in simple local observables as a function of the quench parameter, that is, a non-equilibrium quantum phase transition. We trace the appearance of this non-equilibrium quantum phase transition to an effective metal-insulator transition which occurs on the level of the generalized Gibbs Ensemble (describing the steady-state of the equilibrated system). Studying whether these transitions are robust, we find, in the paradigmatic case of the SSH model, that they persist for both quantum quench protocols of finite duration in time as well as thermal initial states, while they are washed out in the presence of fermion-fermion interactions and for finite system size.
\end{abstract}

\maketitle
\section{Introduction}
Recently, technological advances in the experimental control of ultracold gases \cite{ug1,ug2}, trapped ions \cite{tri}, and nitrogen-vacancy centres in diamonds \cite{nv}, allowed to probe the time evolution of isolated quantum systems. Since, in such systems, the time evolution is unitary, no information about the initial state is lost. However, most often, this information spreads over the whole system, so that, at long times, it is challenging to recollect it. The origin of this behavior lies in the Eigenstate Thermalization Hypothesis (ETH) \cite{eth1,eth2,eth3}, that, qualitatively speaking, states that, in the thermodynamic limit, the expectation value of local observables over any eigenstate with finite energy density can be well approximated by the average over a properly defined thermal density matrix. In this sense, most isolated systems (assuming ETH) thermalize. There are exceptions to this paradigm \cite{violation}. For example, in the Fibonacci chain describing Rydberg atoms\cite{fibo,fibo2}, most eigenstates do follow ETH, while some do not. This phenomenon opened the field of ``quantum many body scars" \cite{scars}. A stronger violation of ETH is provided by many-body systems that have an extensive amount of local or quasi-local conserved quantities. In this case, in fact, the local information stored in the initial wavefunction is preserved by the time evolution. Consequently, systems exhibiting such a behavior can have interesting applications in the field of quantum information \cite{mblquantuminfo}. A first class of systems with quasi-local conserved quantities are many-body localized systems \cite{mbl1,mbl2,mbl3,mbl4,mbl5}. In this case, the quasi-local integrals of motion arise due to real space localization and are robust with respect to weak perturbations. In these systems, there is no simple guideline for building a sensitive effective density matrix for the local observables. A second class of such systems is given by the so-called integrable models \cite{integrable}. In this case, while the violation of ETH is not stable with respect to generic perturbations \cite{thermalgge1,thermalgge2,thermalgge3}, it is indeed possible to build a statistical ensemble capturing the long time expectation value of the local observables. Such an ensemble is called generalized Gibbs Ensemble (GGE) \cite{gge1,gge2,gge3,gge4,gge5,gge6,gge7,gge8,gge9,gge10}. Conceptually speaking it is obtained by maximizing the entropy while taking into account the constraints posed by the local conserved quantities. From the formal point of view, the GGE density matrix could be an exceptionally useful tool, since concepts such as non-equilibrium phase transitions \cite{nept1,nept2,nept3,nept4,nept5,nept6,nept7,nept8,nept9,nept10,nept11} (non-analytical dependencies of long time expectation values as a function of the quench parameter) in integrable systems could be made universal, in this framework, in the very same way transitions in equilibrium are described within the canonical ensemble. However, GGE density matrices are in general difficult to obtain and no systematic link between them and non-equilibrium phase transitions has been performed.\\
\begin{figure}[h!]
	\begin{center}
		\includegraphics[width=\columnwidth]{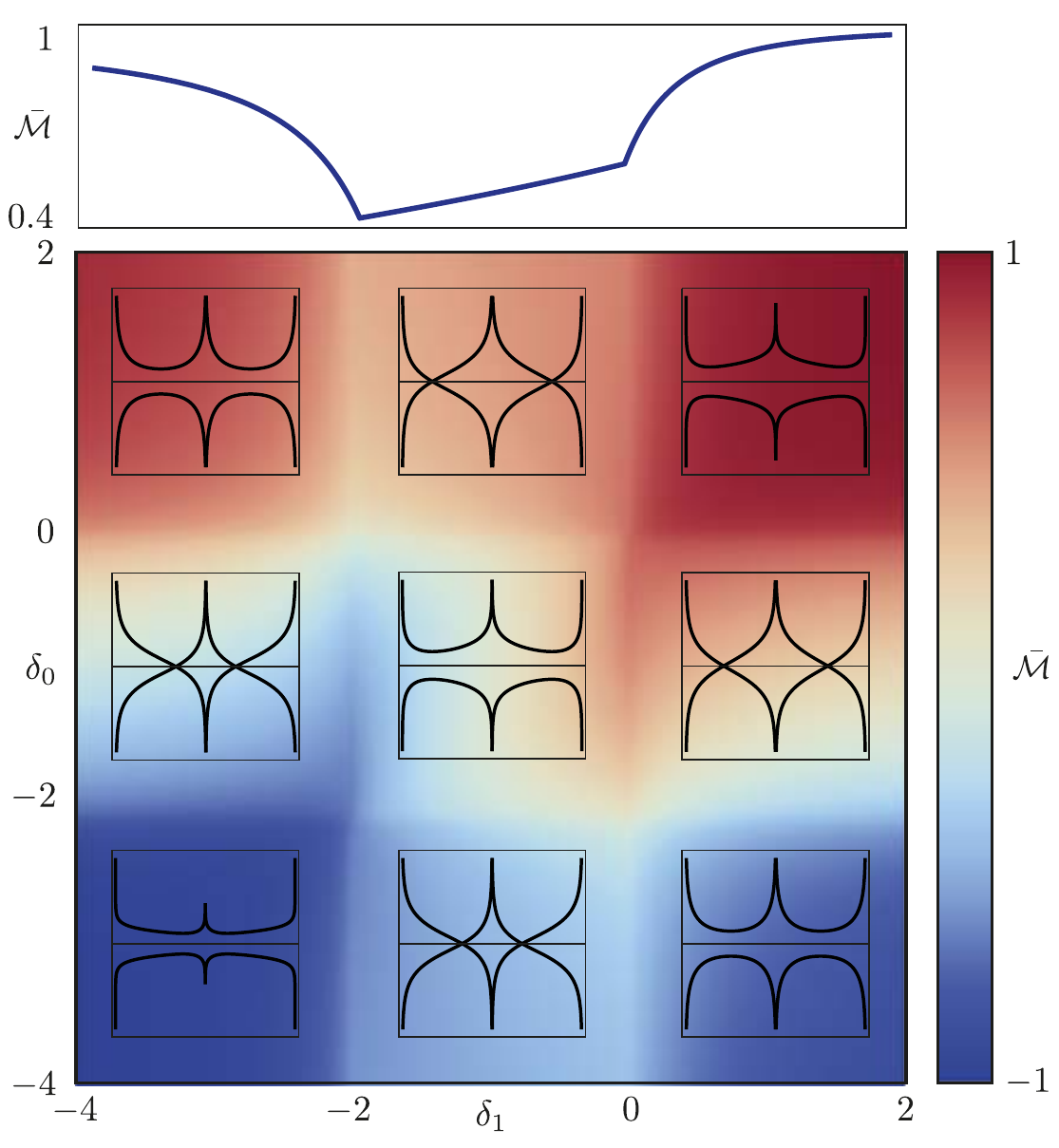}
		\caption{Bottom: Density plot of $\bar{\mathcal{M}}$ as a function of $\delta_0$, $\delta_1$ for a sudden quench and the corresponding typical effective band structure $\xi_{\nu,k}$ (see text). Top: plot of $\bar{\mathcal{M}}$ as a function of $\delta_1$ for $\delta_0=2w$. Here, $\delta_\mu$ is in units $w$.}
		\label{fig:1}
	\end{center}
\end{figure}

In this article, we begin to address this issue in two paradigmatic cases: The Su-Schrieffer-Heeger (SSH) model \cite{ssh1,ssh2}, that represents a starting point for the study of topological phases of matter and the appearance of fractional charges in one dimension\cite{im1,im2,im3,im4,im5,im6,im7,im8,im9,im10,im11}, and the transverse field one-dimensional quantum Ising (QI) model, whose simplicity has opened the way to countless insights in the theory of quantum quenches\cite{calabreseefft,heyl,isi3}. Moreover, the QI model can be mapped onto the Kitaev chain and hence encodes the physics of the so called Majorana bound states\cite{majo}. First, we consider the steady state of a quenched Su-Schrieffer-Heeger (SSH) model. Quenches from an initial Hamiltonian  with a hopping imbalance $\delta_0$ to a final one with $\delta_1$ are considered. In the thermodynamic limit, non-analyticities in observables such as the amount of dimerization $\bar{\mathcal{M}}$ (see below) can occur as a function of $\delta_\mu$ ($\mu=0,1$), signaling {non-equilibrium} quantum phase transitions (QPT). These transitions occur at the same parameter values of $\delta$ where also in equilibrium QPTs (non-analyticities in $\bar{\mathcal{M}}$ in dependence of the hopping imbalance) can be found, but there only at \emph{zero temperature}. In this regard this finding is remarkable, because the quench injects energy into the system~\cite{calabreseefft}, and is in accordance with the findings described in \onlinecite{Moessner}. As we will demonstrate, the GGE density matrix is equivalent to a {grand canonical density matrix} of free fermions, with effective Hamiltonian $\bar{H}$, at finite temperature. Interestingly, tuning the quench parameters, an effective metal-insulator transition (MIT) in $\bar{H}$ is achieved in concurrence with the non-equilibrium QPTs. The generic scenario is summarized in Fig.~\ref{fig:1}, which shows $\bar{\mathcal{M}}$ for a sudden quench $\delta_0\to\delta_1$ and the typical band structure of $\bar{H}$ occurring in each region of the parameters space spanned by the pre-quench  and post-quench value of the gap $\delta_0$ and $\delta_1$, respectively. As we will show, the nature of this effective MIT explains the robustness of the non-equilibrium QPT against the initial preparation of the system and finite-duration quench protocols. The robustness of the non-analytic signatures and the effective MIT render this model a very promising candidate for experimental investigations. We also show that when the model is perturbed in such a way that the GGE does not provide a good description of the long time dynamics, the non-equilibrium QPT is washed out. With this respect, we analyse finite size systems and, in a model which is very similar to the SSH model, non-integrable interaction effects. We then consider quantum quenches in the transverse field QI model. Here, again, the entropy shows kinks as a function of the quench parameter in correspondence to the gapless points. As in the case of the SSH model, this non-equilibrium phase transition occurs together with an effective MIT in the GGE density matrix.\\
Our results suggest that the presence of an effective MIT in the GGE density matrix in connection to an equilibrium QPT leads to a non-equilibrium QPT.\\
The outline of the article is the following. In Sec.II, we inspect the quantum quench dynamics characterizing the SSH model and, in Sec.III, we discuss the same physics in the context of the transverse field QI model. Finally, in Sec.IV, we draw our conclusions.
\section{Quantum quench in the SSH model}
\subsection{Sudden quench}
The momentum space Hamiltonian for the quenched SSH model (on a finite ring of length $L$ with $\mathcal{N}$ unit cells and lattice constant set to one in the following) is given by \cite{ssh2} $H(t)=\sum_k \Psi^\dag_k \left\{\sigma_x \left[w+w\cos (k)+\delta(t)\right]+w\sigma_y \sin(k)\right\}\Psi_k$. Here, $\Psi^\dagger_k=(c^\dagger_{k,A},c^\dagger_{k,B})$ is a Fermi spinor, $A$ and $B$ the two sublattice labels and $k=2\pi j/\mathcal{N}$ with $|j|\leq\mathcal{N}$. Furthermore, $\sigma_{i}$ are Pauli matrices and $w$ is the hopping energy. The hopping imbalance term $\delta(t)$, which in equilibrium determines the gap, encodes the quench details: In most of the paper a sudden change $\delta(t)=\delta_0\theta(-t)+\delta_1\theta(t)$ - with $\theta(t)$ the Heaviside step function - is considered. The SSH Hamiltonian is diagonalized as
\begin{equation}
\label{eq:Hlat}
H_{\mu}=\sum_{k}\epsilon_{\mu,k}\left[d^{\dagger}_{\mu,c,k}d_{\mu,c,k}-d^{\dagger}_{\mu,v,k}d_{\mu,v,k}\right]
\end{equation}
with $\mu=0$ ($\mu=1$) for $t<0$ ($t>0$), $d_{\mu,\nu,k}$ fermionic operators for the $\nu=c,v$ bands and $\epsilon_{\mu,k}=\sqrt{\delta_\mu^2+2(w^2+w\delta_{\mu})[1+\cos(k)]}$. In the initial state ($t<0$) the system is prepared in the ground state $|G_0\rangle$ of $H_0$.\\
\noindent In the thermodynamic limit, the quantum average of local observables $\mathcal{O}(t)=\langle G_0| O(t)|G_0\rangle$ approaches a steady value $\bar{\mathcal{O}}=\mathcal O(t\to\infty)$ with a typical $\propto t^{-1}$ power-law decay (not shown). Since the system is integrable, this steady value can also be obtained as the trace $\bar{\mathcal{O}}=\langle\mathcal{O}\rangle\equiv\mathrm{Tr}\{O(0)\rho_{G}\}$ over the GGE density matrix \cite{gge1} constructed via the conserved charges $N_{\nu,k}=d^{\dagger}_{1,\nu,k}d_{1,\nu,k}$ as
\begin{equation}
\label{eq:rhog}
\!\!\!\rho_{G}=\frac{e^{-\sum_{\nu,k}\lambda_{\nu,k}N_{\nu,k}}}{Z_{G}} , Z_{G}=\mathrm{Tr}\left\{e^{-\sum_{\nu,k}\lambda_{\nu,k}N_{\nu,k}}\right\},
\end{equation}
where $\lambda_{\nu,k}=\log(n_{\bar{\nu},k}/n_{\nu,k})$, with $n_{\nu,k}=\langle G_0|N_{\nu,k}|G_0\rangle$ and $\bar{\nu}=v/c$ if $\nu=c/v$.

A physically interesting way of building the Lagrange multipliers giving the GGE density matrix in the case of sudden quench from the ground state is by the transformation $\mathcal{U}_{0,k}^{1}=e^{i\vec{\mathcal{D}}_k\cdot\vec{\sigma}}$ connecting post-quench Fermi operators $d_{1,\nu,k}$ to the pre-quench ones $d_{0,\nu,k}$, with $\vec{\sigma}$ the vector of Pauli matrices. The norm $|\vec{\mathcal{D}}_k|=\arctan\left\{\frac{\sqrt{4-(1-\Delta_k)^2}}{1-\Delta_k}\right\}$ plays a central role in defining the behavior of the quench. In fact, the function $ \Delta_{k}$ is such that~(cf \ref{SM:eq:Delta},~\ref{SM:eq:effbands})
\begin{equation}
\label{eq:popl}
n_{c/v,k}=\frac{1\pm\Delta_k}{2} , \quad \lambda_{c/v,k}=\pm\log\left(\frac{1-\Delta_k}{1+\Delta_k}\right),
\end{equation}
and thus directly controls the GGE. One finds $|\Delta_k|\leq 1$ with $0\leq|\vec{\mathcal{D}}_k|\leq\pi/2$. Furthermore, $|\Delta_k|=1$ {\em only} for $k=0,\pm\pi$: in particular one has
\begin{equation}
\label{eq:FP}
\Delta_0=-s(\delta_0+2)s(\delta_1+2) ,\, \Delta_{\pm\pi}=-s(\delta_0)s(\delta_1)\,,
\end{equation}
with $s(x)=|x|/x$ the sign function. When $\Delta_{k}=-1$, the transformation reduces to the identity $\mathcal{U}_{0,k}^{1}=\sigma_0$ and the quench does not affect the populations, while for $\Delta_{k}=1$ the transformation $\mathcal{U}_{0,{k}}^{1}=i\sigma_y$ induces a {swap} of the $c,v$ states. The value of $\Delta_k$ at $k=0,\pm\pi$ is constant and insensitive to variations of the quench parameters provided they remain within one of the nine regions bounded by the lines $\delta_\mu=0,-2w$ (see lower panel of Fig.~\ref{fig:1}). On the other hand, crossing one of the boundaries results in a sharp, non-analytical jump in  $\Delta_k$. Thus, the center and edges of the Brillouin zone (BZ) act for the quench as fixed points, whose character is determined by the quench parameters. A detailed study of $ \Delta_k $~(see Appendix \ref{SM:sec:Delta}) allows us to identify four non-contiguous regions out of the nine defined above: Here, when $ \Delta_0\Delta_\pi <0 $, a non-trivial {\em inversion of population}~\cite{FN0}, characterized by $n_{\nu,0}n_{\bar{\nu},\pm\pi}=0$, takes place.

As mentioned, the transformation is directly linked to the GGE, since, in the sudden quench case from the ground state, one has $n_{\nu,k}^{-1}={1+e^{\xi_{\nu,k}\beta^{*}}}$
where $\xi_{\nu,k}=w\lambda_{\nu,k}$ and $\beta^{*}=w^{-1}$ is an inverse temperature. One can hence exactly rephrase the GGE density matrix as the  Grancanonical ensemble of free fermions with Hamiltonian 
\begin{equation}\bar{H}=\sum_{\nu,k}\xi_{\nu,k}d^\dag_{1,\nu,k}d_{1,\nu,k},
\end{equation}
inverse temperature $\beta^{*}$, and zero chemical potential. Combining the above analysis and Eq.~(\ref{eq:popl}) one can conclude that when a non-trivial inversion of population of the bands $\epsilon_{\nu,k}$ of the post-quench Hamiltonian is present, the effective bands $\xi_{\nu,k}$ cross zero energy and thus have a {metallic} character, while no crossing occurs in all other cases and the bands $\xi_{\nu,k}$ have an {insulating} character. When quench parameters cross one of the boundary lines described above, thus an effective MIT in $\bar{H}$ occurs. This is exemplified as black graphs in the nine different tiles of Fig.~\ref{fig:1} separated by $\delta_\mu=0,-2w$.  We stress here that the MIT is an effective one showing up in the GGE. How this reflects in physically relevant (local) observables is a priori unclear, but in our case we will show explicitly in the following that its imprint is quite pronounced.

Next we analyse how the effective MIT influences observables of interest. The most intuitive one to investigate is the average level of dimerization $\bar{\mathcal{M}}$, given by the expectation value of $\mathcal{M}(x)=\Psi_{x}^{\dagger}\sigma_{x}\Psi_{x}$ with $\Psi_x=\sum_{k}e^{ikx}\Psi_{k}/\sqrt{L}$. Note that translational invariance implies that the expectation value is independent of the position. The main panel of Fig.~\ref{fig:1} shows a density plot of $\bar{\mathcal{M}}$ as a function of $\delta_\mu$. Crossing any of the transition lines $\delta_\mu=0,-2w$, a kink in $\bar{\mathcal{M}}$ is encountered. The top panel shows results for $\delta_0=2w>0$: the discontinuity in $\partial_{\delta_1}\bar{\mathcal{M}}$ at $\delta_1=0,-2w$ is evident. These kinks represent a signature of the occurrence of the effective MIT. Their origin is the non-analytic dependencies of the populations at $k=0,\pm\pi$ combined with the fact that, in the thermodynamic limit, the density of states of such points diverges as the curvature of $\epsilon_{1,k}$ vanishes at these points. Several other quantities show a similar behavior.

\begin{figure}[htbp]
	\begin{center}
		\includegraphics[width=\columnwidth]{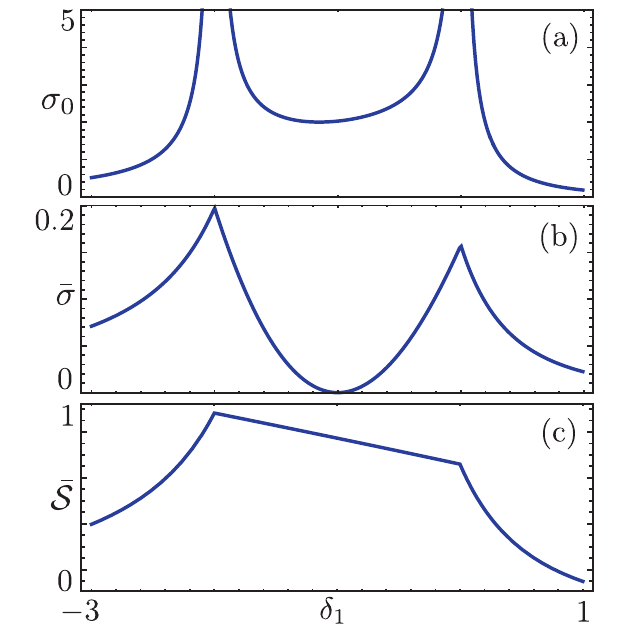}
		\caption{(a) Plot of $\bar{\sigma}_0$ (units $w^2$); (b) Plot of $\bar{\sigma}$ (units $w^2$) ; (c) Plot of $\mathcal{S}$ as a function of $\delta_1$ (units $w$) for $\delta_0=2w$.}
		\label{fig:2}
	\end{center}
\end{figure}

Given the presence of the effective MIT in the GGE, we inspect the fluctuation of the space-averaged effective ``current" $J_0=\sum_{\nu,k}\left(\partial_{k}\xi_{\nu,k}\right)N_{\nu,k}$. Such fluctuation is defined using the phase velocity of the effective bands~\cite{mahan,FN2}, and in the steady state limit one has $\bar{\sigma}_0=\langle J_0^2\rangle$~(see Appendix \ref{SM:subsec:CurrFluc}). This quantity is shown in Fig.~\ref{fig:2}(a) for $\delta_0>0$. For $\delta_1=0$ (and $\delta_1=-2w$) it diverges $\propto|\delta_1|^{-1}$ (and $\propto|\delta_1+2|^{-1}$). Furthermore, fluctuations are larger in the effective metallic phase, while they tend to vanish in the insulating one, as one would expect \cite{mahan}. Although $\bar{\sigma}_0$ is not a directly accessible quantity, signatures of the effective MIT are present also in the steady state fluctuations $\bar{\sigma}=\langle J^2\rangle $ of the space-averaged physical current $J=\sum_{k}\left(\partial_{k}\epsilon_{\nu,k}\right)N_{\nu,k}$, shown in Fig.~\ref{fig:2}(b). In contrast to the  current fluctuations in the effective picture though, here no marked differences in the magnitudes are found in the different phases. However, kinks occur at the boundaries between the phases. As a third example, Fig.~\ref{fig:2}(c) shows the thermodynamic entropy~(see Appendix \ref{SM:subsec:Entropy}) $\bar{\mathcal{S}}$ of the system for $\delta_0>0$: It is largest in the metallic phase and displays kinks for $\delta_1=0,-2w$. This quantity is particularly interesting, since it is intrinsic to thermodynamics.\\

\subsection{Robustness} As shown above, signatures of the effective MIT occur in a vast array of quantities. It is important to establish how robust the results are. We will consider $\bar{\mathcal{M}}$ as an example but the conclusions drawn below apply to all quantities discussed above.
\begin{figure}[htbp]
	\begin{center}
		\includegraphics[width=\columnwidth]{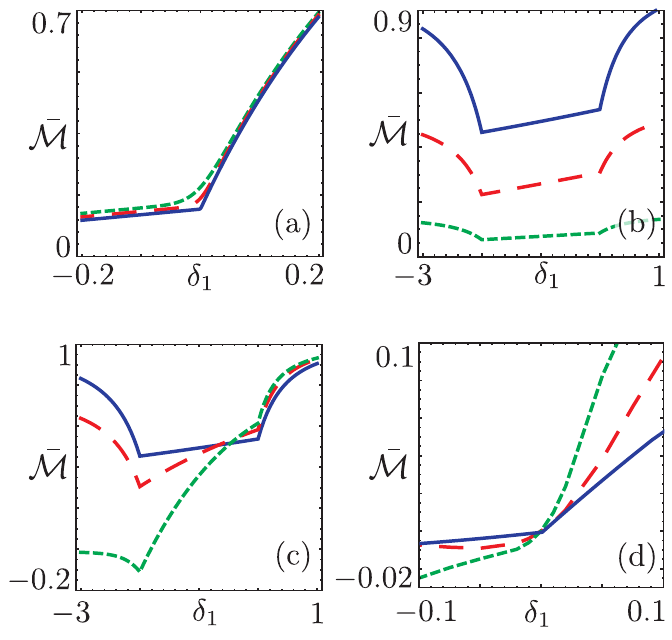}
		\caption{Plot of $\bar{\mathcal{M}}$ as a function of $\delta_1$ (in units of $w$) for different (a) number of lattice sites: solid the thermodynamic limit, dashed $\mathcal{N}=80$, dotted $\mathcal{N}=20$. The last two curves are averaged over a period - see text; (b) temperature of the initial state: solid $T=0$, dashed $T=5$, dotted $T=20$; (c) duration of the quench ramp: solid $\tau=0$, dashed $\tau=2$, dotted $\tau=10$; (d) Strength of the fermion-fermion interaction - see text: solid $U=0$, dashed $U=1$, dotted $U=2$, $T=0$. Here, $\delta_0=5w$ in panels (a-c) and $\delta_0=w$ in panel (d), $T$ is in units of $wk_B^{-1}$ with $k_B$ the Boltzmann constant, $\tau$ is in units of $\hbar w^{-1}$ and $U$ is in units of $w$.}
		\label{fig:3}
	\end{center}
\end{figure}
We begin by discussing deviations from the thermodynamic limit. With a finite number of lattice sites, averages do not converge to a steady value for long time but oscillate with a finite recursion time: the GGE hypothesis fails altogether. Figure~\ref{fig:3}(a) shows the quantum and time average over a period of $\mathcal{M}(t)$ near $\delta_1=0$. Dashed and dotted curves, calculated with a finite number of lattice sites, show that kinks are smoothed out as the number of sites decreases. This confirms the thermodynamic limit as a crucial ingredient for the non-analyticities to arise in the GGE predictions. Interestingly, though, even for $\mathcal{N}$ as small as $20$ one can still observe a distinct imprint of the non-analyticities  found for $N\to\infty$ in the time-averaged $\bar{\mathcal{M}}$.  

Furthermore, the features are robust in the case of a {thermal} preparation of the initial state~\cite{FN3}. Figure~\ref{fig:3}(b) shows $\bar{\mathcal{M}}$ obtained for an initial state at different temperatures $T$: although the curves are quantitatively different, with a global suppression of the dimerization, non-analyticities are always present. The origin of the robustness is that, for an initial temperature $T$, one has $n_{c,k}-n_{v,k}=f_{T,k}\Delta_k$ with $f_{T,k}=\sinh(\epsilon_{0,k}/k_BT)/[1+\cosh(\epsilon_{0,k}/k_BT)]>0$~(cf \ref{SM:eq:thermal}). This result means that the effective MIT occurs in the same parameter regions as in the $T=0$ case~\cite{FN4}. Note that the robustness of the non-analyticity with respect to temperature is particularly intriguing since it is not present in the equilibrium QPT characterizing the model.

We then consider the  case of a quantum quench of finite time duration $\tau$, where the quench protocol is described by a linear ramp. Typical results are shown in Fig.~\ref{fig:3}(c). The non-analytic behavior persists, although results again differ quantitatively. This is due to the robustness of the effective MIT, that can be demonstrated by showing that the fixed points of the quench transformation only differ by an additional phase shift with respect to the case of sudden quench~(see Appendix \ref{SM:sec:Delta}).

Finally, we address the effects of static inter-particle interactions. We consider here a very similar model which - in the absence of interactions - displays the same qualitative behavior than the one discussed so far, but it is easier to simulate. The Hamiltonian is given by $H(t)=\sum_i^{N} w c_i^\dagger c_{i+1}+{\rm H.c.}+\delta(t)(-1)^i n_i+Un_{i}n_{i+1}$ and we consider $\bar{\mathcal{M}}=\left\langle n_0-1/2\right\rangle$ as an observable. $c_i^{(\dagger)}$ annihilates (creates) a spinless fermion on lattice site $i$. The model thus describes spinless fermions on a one-dimensional chain with staggered field $\delta(t)$ and nearest-neighbor interaction $U$. At time $t=0$ the staggered field is subject to the quench $\delta(t)=\delta_0\theta(-t)+\delta_1\theta(t)$, abruptly changing its value from $\delta_0$ to $\delta_1$.  This model can be simulated with relative ease using standard density matrix renormalization group techniques based on matrix product states \cite{White92,Schollwock11,Kennes16}. The time scales which can be reached are bound within this approach by the entanglement growth of the system and, thus, the steady state behavior has to be read off at large but finite times. For $U=0$ strong oscillations in the dynamics after the quench render such an extrapolation difficult, but for this particular parameter value exact methods can be employed to extract the asymptotic behavior. At finite $U$ these oscillations are strongly damped out allowing for a straightforward extrapolation to long times~(see Appendix \ref{SM:sec:Interactions}).   The inclusion of the interaction term makes the model non-integrable, which in turn is believed to destroy the GGE picture. Fig.~\ref{fig:3}(d) shows results for different values of the interaction strength: non-analyticities are washed out, as would be expected, by a thermal redistribution of the excitation energy in the long-time limit.\\
\section{Quantum quench in the Ising Model}
\begin{figure}[htbp]
	\begin{center}
		\includegraphics[width=\columnwidth]{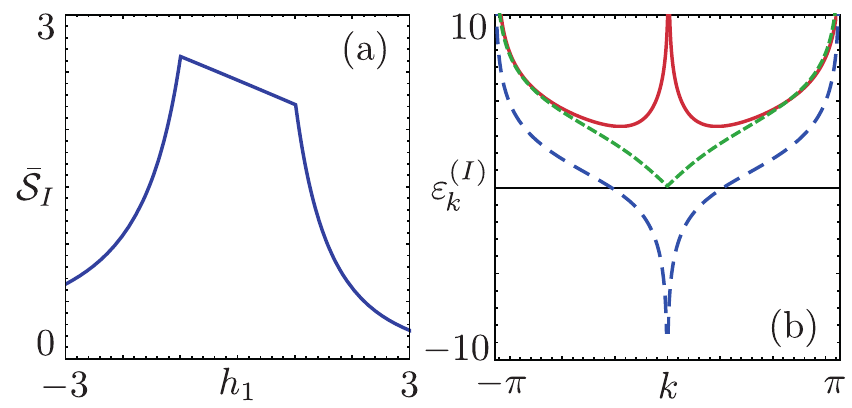}
		\caption{(a) Plot of $\mathcal{\bar{S}}_I$ as a function of $h_1$, for $ h_0=0 $; (b) Plot of ${\varepsilon^{(I)}_k}$, as a function of $k$, for $h_0=2$ and $h_1=5$ (red solid line), $h_1=1$ (green dashed line), and $h_1=0$ (blue dashed line). The thin black line corresponds to the chemical potential.}
		\label{fig:4}
	\end{center}
\end{figure}

In this section, we consider the transverse field QI model. The Hamiltonian is
\begin{equation}
H_I(t)=-\sum_{j=1}^\mathcal{N}\frac{1}{2}\left[\sigma^x_{j+1}\sigma^x_j+h(t)\sigma^z_j\right]
\label{eq:c}\end{equation}
where $\sigma^\alpha_j$, $\alpha = x, y, z$, are the Pauli matrices at site $j$ of a chain of $\mathcal{N}$ sites with periodic boundary conditions, and $h(t)$ is the transverse field. We consider sudden quantum quenches, so that $h(t)=h_0\theta(-t)+h_1\theta(t)$, and we impose the system to be in the ground state $|0_I\rangle$ for $t<0$, and to evolve unitarily for $t>0$. Note that the state $|0_I\rangle$ is uniquely defined, even in the thermodynamic limit, since we consider $h_0>1$.\\
The Hamiltonian $H_I(t)$ can be diagonalized at any time by means of a Wigner-Jordan transformation onto spinless fermions, followed by a Bogoliubov transformation\cite{isi3}. In the even parity sector, relevant for the case inspected since we perform a quantum quench from the ground state at $h_0>1$, the diagonal forms of the pre($t<0$)/post$(t>0)$ quench Hamiltonians $H_I^{(i)}$ ($i=0/1$ respectively) read as
\begin{equation}
H_I^{(i)}=\sum_{k=-N}^{N-1}\xi^{(i)}_k
\left( b^{(i)\dagger}_kb^{(i)}_k-\frac{1}{2}\right),
\label{eq:cc}\end{equation}
with
\begin{equation}
\xi^{(i)}_k=\sqrt{[h_i-\cos (p_k)]^2+\sin^2(p_k)}.
\end{equation}
Here, $p_k=2\pi k/\mathcal{N}$ and $b^{(i)}_k$ are fermionic operators. Note that $b^{(0)}_k|0_I\rangle=0$, for every $k$. For the details of the transformation rewriting Eq.~\ref{eq:c} to Eq.~\ref{eq:cc}, see, for example, Ref.~\onlinecite{franchini}. The fermionic occupation numbers $N^{(I)}_k=b^{(1)\dagger}_kb^{(1)}_k$ and their averages ${n}^{(I)}_k=\langle 0_I|{N}^{(I)}_k|0_I\rangle$ allow to define, in the thermodynamic limit and for times $t\rightarrow\infty$, the post quench thermodynamic entropy $\mathcal{\bar{S}}_I=-\sum_k {n}^{(I)}_k\ln ({n}^{(I)}_k)+(1-{n}^{(I)}_k)\ln(1-{n}^{(I)}_k)$ and the GGE density matrix of the system. The latter quantity reads as 
\begin{equation}
\rho_{G}^{(I)}=\frac{e^{-\sum_k \varepsilon^{(I)}_k N^{(I)}_k}}{Z_G^{(I)}},{Z_G^{(I)}}=\mathrm{Tr}\left\{e^{-\sum_k \varepsilon^{(I)}_k N^{(I)}_k}\right\},
\end{equation}
with $\varepsilon^{(I)}_k$ implicitly given by
\begin{equation}
{n}^{(I)}_k=\frac{1}{e^{\varepsilon^{(I)}_k}+1}.
\end{equation}
Again, we can interpret the GGE density matrix as a Grancanonical density matrix, at temperature set to unity and at zero chemical potential, for fermions with effective Hamiltonian
\begin{equation}
\bar{H}^{(I)}=\sum_k {\varepsilon^{(I)}_k}b^{(1)\dagger}_kb^{(1)}_k.
\end{equation}
As in the case of the SSH model, the entropy shows kinks as a function of the quench parameter, in correspondence to the gapless points of the dispersion relation signalling the equilibrium QPT between the paramagnetic and the ferromagnetic phase. Correspondingly the effective Hamiltonian $\bar{H}^{(I)}$ undergoes a metal insulator transition. The analogy to the behavior in the SSH model is hence complete. Examples are given in Fig.~\ref{fig:4}. In panel (a), the entropy $\mathcal{\bar{S}}_I$ is plotted as a function of $h_1$, for $h_0=10$. $\mathcal{\bar{S}}_I$ is shown to have non-analyticities in correspondence to the equilibrium QPTs occurring at $h_1=\pm 1$. In panel (b), the effective energies ${\varepsilon^{(I)}_k}$ are plotted, as a function of $k$, for $h_0=2$ and $h_1=5$ (red solid line), $h_1=1$ (green dashed line), and $h_1=0$ (blue dashed line). As in the case of the SSH model, these effective bands undergo an effective MIT in correspondence to the equilibrium QPT. In fact, for $h_1>1$ the dispersion does not cross the chemical potential (zero in this case), while for $h_1<1$ it does.
\section{Conclusions}
While in a previous work\cite{Moessner} the highly non-trivial relation between equilibrium and non-equilibrium QPT was inspected with reference to the topological nature of the equilibrium QPT, we have here adopted a different perspective, more suitable to generalizations in the context of integrable systems. We have observed that, in the paradigmatic cases of the SSH model and of the transverse field QI model, the non-equilibrium QPTs appear in connection to both an equilibrium QPT and an effective MIT in the GGE density matrix of the system. By direct inspection in the case of the SSH model, we have also shown that the non-equilibrium QPT is indeed robust with respect to those perturbations that do not spoil the validity of the GGE, and hence the presence of the effective MIT.

The phenomenology we describe appears general and should hold true also for higher dimensional systems. An interesting extension to our work includes the discussion of terms that break integrability only weakly. The results we report should carry over to the prethermal state reached in these situations. 

\textit{Acknowledgements---}
D.M.K. acknowledges support by the Deutsche Forschungsgemeinschaft through the Emmy Noether program (KA 3360/2-1). Simulations were performed with computing resources granted by RWTH Aachen University under projects rwth0013 and prep0010. N.T.Z. acknowledges financial support by the DFG (SPP1666 and SFB1170 ToCoTronics),the Helmholtz Foundation (VITI), the ENB Graduate school on Topological Insulators. Interesting discussions with C. Fleckenstein, D. Hetterich, M. Serbyn, C. P{\'e}rez-Espigares and B. Trauzettel are also acknowledged. 

\appendix
\section{Quantum quench and geometrical interpretation}\label{SM:sec:Delta}
\subsection{Quench-induced transformation in the SSH model}\label{SM:subsec:transf}
We start this section by giving, as stated in the main text, the Hamiltonian of the Su-Schrieffer-Heeger (SSH) model~\cite{ssh2,ssh1}
\begin{equation}\label{SM:eq:H(t)}
H(t)=\sum_k \Psi^\dagger_k \left\{\sigma_x\left[w+w\cos k+\delta(t)\right]+w\sigma_y \sin k\right\} \Psi_k ,
\end{equation}
where  $ \Psi_k^\dagger= \left(c_{k,A}^\dag,c_{k,B}^\dag \right) $ is a two-component momentum resolved Fermi spinor, $ A $ and $ B $ represent the two sublattices of the unit cell and the hopping between the same and different cells is staggered. This difference is encoded in the quantity $ \delta(t) $, which measures the amplitude of the gap in the spectrum of the system. In the context of sudden quantum quenches, $ \delta(t) $ abruptly changes its value, namely
\begin{equation}
\delta(t) = \delta_0 \theta(-t)+\delta_1 \theta(t) .
\end{equation}
Here, we conveniently use the indexes 0 or 1 for the pre- or post-quench quantities, respectively, and the symbol $ \theta $ denotes the Heaviside step function. The Hamiltonian, accordingly, can be written as 
\begin{equation}
H(t)= H_0 \theta(-t)+H_1 \theta(t) .
\end{equation}
It is useful, at this point, to diagonalize both the pre- and post-quench Hamiltonians, by means of a unitary transformation, to obtain ($ \mu=0,1 $)
\begin{equation}
H_\mu= \sum_k \epsilon_{\mu,k}\left( d^\dagger_{\mu,c,k} d_{\mu,c,k}- d^\dagger_{\mu,v,k} d_{\mu,v,k}\right),
\end{equation}
where the subscripts $c$ and $v$ are associated with the conduction and valence bands respectively, and
\begin{equation}
\epsilon_{\mu,k}=\sqrt{\delta_\mu^2+2(w^2+w\delta_\mu)(1+\cos k)}
\end{equation} 
is the energy spectrum. The transformation, which relates the lattice and the diagonal bases, is defined as:
\begin{equation}
\label{eq:S5}
\Phi_{\mu,k}=\left(\begin{array}{c}
d_{\mu,c,k}\\ 
d_{\mu,v,k} 
\end{array}
\right) = U_{\mu,k} \Psi_k ,
\end{equation}
where 
\begin{equation}
U_{\mu,k} =\left(\begin{array}{cc}
A_{\mu,k} & B_{\mu,k} \\ 
-B_{\mu,k}^* & A_{\mu,k}
\end{array} \right)
\end{equation}
and 
\begin{equation}
\label{eq:S7}
A_{\mu,k}=\dfrac{1}{\sqrt{2}}\ ,\qquad B_{\mu,k}=\dfrac{1}{\sqrt{2}}\dfrac{w\left(1+e^{-i k}\right)+\delta_\mu}{\epsilon_{\mu,k}}\ .
\end{equation}
One can easily compose these transformations to get the unitary matrix which connects the two diagonal pre- and post-quench bases. It has the following form:
\begin{equation}\label{SM:eq:Transfbases}
\Phi_{1,k}= U_{1,k} U_{0,k}^\dag \Phi_{0,k} =\mathcal{U}_{0,k}^1 \Phi_{0,k} .
\end{equation} 
\subsection{General properties}\label{SM:subsec:genprop}
To get more insight about the quench-induced transformation, it is instructive to rewrite $ \mathcal{U}_{1,k}^0 $ in the following form,
\begin{equation}
\mathcal{U}_{1,k}^0 = \exp\left(i \vec{\mathcal{D}}_k\cdot \vec\sigma\right)
\end{equation}
where $ \vec{\sigma} $ is the vector of Pauli matrices and $ \vec{\mathcal{D}}_k = |\vec{\mathcal{D}}_k|\ \vec n_k$. By exploiting the properties of Pauli matrices one obtains (from here on we set $ w=1$ for simplicity)
\begin{equation}
|\vec{\mathcal{D}}_k|= \arctan\left[\dfrac{\sqrt{4-(1-\Delta_k)^2}}{1-\Delta_k}\right] ,
\end{equation}
\begin{equation}
\vec n_k= \dfrac{1}{\sqrt{4-(1-\Delta_k)^2}}\left(\begin{array}{c}
-\sin k \left(\dfrac{1}{\epsilon_{1,k}}+\dfrac{1}{\epsilon_{0,k}}\right) \\ \dfrac{1+\delta_1+\cos k}{\epsilon_{1,k}}-\dfrac{1+\delta_0+\cos k}{\epsilon_{0,k}}
\\ \dfrac{\delta_0-\delta_1}{\epsilon_{0,k}\epsilon_{1,k}}\sin k
\end{array} \right) ,
\end{equation}
\begin{equation}
\Delta_k =- \dfrac{1+(1+\delta_0)(1+\delta_1)+(2+\delta_0+\delta_1)\cos k}{\epsilon_{0,k}\epsilon_{1,k}} .
\end{equation}
The function $ \Delta_k $ introduced above emerges naturally from the transformation, i.e. from the sudden quench: note that, indeed, $|\vec{\mathcal{D}}_k|$ only depends on the momentum via $ \Delta_k $. The following general properties hold:
\begin{enumerate}
	\item $ \Delta_k $ is a periodic and analytic function {\em of the momentum} $ k $, with $ \Delta_k =  \Delta_{-k} $ due to time-reversal symmetry;
	\item $-1\leq\Delta_k\leq 1$, which in turns implies ${0\leq|\vec{\mathcal{D}}_k|\leq\pi/2}$.
\end{enumerate}
Moreover, the unit vector $ \vec{n}_k $ shows a clear symmetry with respect to the parameter $ k $, such that for $ k\rightarrow -k $ the vector $ \vec{\mathcal{D}}_k $ gets mirrored about the $ y $-axis. In view of the properties outlined above let us study $\Delta_k$ over half of the BZ, namely on $\mathcal I=[0,\pi]$, where one finds that the equation $|\Delta_k|=1$ has solutions only if $k=0,\pi$. More specifically, denoting $s(x)=x/|x|$ the sign function, one finds
\begin{equation}
\Delta_0=-s(\delta_0+2)s(\delta_1+2)\ ;\ \Delta_{\pi}=-s(\delta_0)s(\delta_1).
\end{equation}
Clearly, as a function of the quench parameters, $\Delta_k$ is {\em not} analytic but exhibits {\em jumps} when the critical lines $\delta_\mu=0$ and $\delta_{\mu}=-2$ are crossed, while for all other values of $k$, $\Delta_k$ is instead a continous and analytic function of the quench parameters. Note that at $\delta_\mu=0,-2$ the equilibrium SSH model presents two quantum critical points (QCPs) associated to a quantum phase transition (QPT). This defines nine regions in the $(\delta_0,\delta_1)$-plane: within each region the values of $\Delta_0$ and $\Delta_\pi$ are constant and independent of the quench. One finds
\begin{align}
\Delta_k=-1\implies\mathcal{U}_{1,k}^0  =\left(\begin{array}{cc}
1 & 0 \\ 
0 & 1
\end{array} \right) ,\\ \Delta_k=1\implies\mathcal{U}_{1,k}^0  =\left(\begin{array}{cc}
0 & 1 \\ 
-1 & 0
\end{array} \right).
\end{align}
When $\Delta_k=-1$ the $c,v$ states are unchanged, while for $\Delta_k=1$ the $c,v$ states are essentially swapped. In addition, $\Delta_k=1\implies\vec{n}_k=(0,-1,0)$. Therefore, the transformation $\mathcal{U}_{1,k}^0$ has two fixed points, namely $\vec{D}_k=(0,0,0)$ (henceforth called I) where it reduces to the identity and $\vec{D}_k=(0,-\pi/2,0)$ (henceforth called R), where bands are swapped. 
\subsection{Occupation numbers and GGE weights}\label{SM:subsec:GGE}
In this section we introduce the Generalized Gibbs Ensemble (GGE)~\cite{gge10} which, in the thermodynamic limit, reproduces the long time limit of the expectation value of the system observables. We start by presenting the GGE density matrix, obtained by maximizing the entropy while keeping into account the conservation of the occupation number operators $N_{\alpha,k}=d^{\dagger}_{1,\alpha,k}d_{1,\alpha,k} $,
\begin{equation}
\rho_G=\dfrac{e^{-\sum_{\alpha,k}\lambda_{\alpha,k}N_{\alpha,k}}}{\tr\left\{e^{-\sum_{\alpha,k}\lambda_{\alpha,k}N_{\alpha,k}}\right\}} ,
\end{equation}
where $ \alpha=c,v$ and $ \lambda_{\alpha,k} $ are the corresponding Lagrange multipliers, obtained by imposing
\begin{equation}
\tr\left\{N_{\alpha,k} \rho_G \right\} = \langle G_0|N_{\alpha,k}|G_0\rangle =n_{\alpha,k} ,
\end{equation}
with $|G_0\rangle$ the pre-quench ground state. One has
\begin{align}\label{SM:eq:ncnv}
\langle G_0|N_{c,k}|G_0\rangle=\left|A_{0,k}B_{1,k}-A_{1,k}B_{0,k}\right|^2, \\ \langle G_0|N_{v,k}|G_0\rangle=\left|A_{0,k}A_{1,k}+B_{1,k}^*B_{0,k}\right|^2 ,
\end{align}
and
\begin{equation}
\lambda_{c,k}=\ln\left(\dfrac{n_{v,k}}{n_{c,k}}\right)=-\lambda_{v,k} .
\end{equation}
Interestingly, by recalling Eq.~\eqref{SM:eq:ncnv}, we observe that
\begin{equation}\label{SM:eq:Delta}
\Delta_k=n_{c,k}-n_{v,k} ,
\end{equation}
and
\begin{equation}\label{SM:eq:effbands}
\lambda_{c,k}= \ln\left(\dfrac{1-\Delta_k}{1+\Delta_k}\right).
\end{equation}
Equation~(\ref{SM:eq:Delta}) is particularly interesting since it links the imbalance between $c$ and $v$ states to the function $\Delta_k$, which, in turn, is directly connected with the presence of the quench. With $|G_0\rangle$ as the pre-quench state, $\Delta_k=1$ implies a complete inversion of population. Three different scenarios can occur, according to the quench parameters:
\begin{itemize}
	\item If $\Delta_0=-1$ and $\Delta_\pi=-1$, the function $\Delta_k$ must have at least one maximum in each half of the BZ. It is easy to prove that in this situation $\Delta_k<0$ always. As a result, no inversion of population occurs;
	\item If $\Delta_0=\mp 1$ and $\Delta_\pi=\pm 1$, the function $\Delta_k$ must have at least one zero in each half of the BZ. Indeed, there is exactly one zero per half, located at
	\begin{equation}
	k^{*}=-\arccos\left(\dfrac{2+\delta_0+\delta_1+\delta_0\delta_1}{2+\delta_0+\delta_1}\right)\,.
	\end{equation}
	In this situation a non-trivial inversion of population occurs for $-\pi\leq k<-k^{*}$ and $k^{*}<k\leq\pi$;
	\item If $\Delta_0=1$ and $\Delta_\pi=1$, the function $\Delta_k$ must have at least one minimum in each half of the BZ and one can prove that $\Delta_k>0$ always. As a result, a complete inversion of population in the whole BZ occurs. Formally, this last case can be analyzed by simply swapping the role of the $c$ and $v$ states throughout the entire BZ, thus we label this a ``trivial" inversion of population.
\end{itemize}
\subsection{Effective GGE energy bands}\label{SM:subsec:geom}
Upon defining $\varepsilon_{\alpha,k}=w\lambda_{\alpha,k}$ and introducing a fictitious effective inverse temperature $\beta^{*}=w^{-1}$ we can re-write
\begin{equation}
n_{\alpha,k}=\dfrac{1}{1+e^{\beta^{*}\varepsilon_{\alpha,k}}}\,.
\end{equation}
Thus, the occupation numbers $n_{\alpha,k}$ correspond to a thermal distribution of free fermions with effective energy bands $\varepsilon_{\alpha,k}$ and zero chemical potential. By exploiting this analogy, from the above discussion and Eq.~(\ref{SM:eq:effbands}), we can conclude that:
\begin{itemize}
	\item If $\Delta_0\Delta_{\pi}=1$, the two effective bands never touch nor cross the chemical potential and thus describe an effective insulating configuration;
	\item If $\Delta_0\Delta_{\pi}=-1$, the two effective bands cross precisely at chemical potential, exactly once per half of the BZ, and thus describe an effective metallic configuration.
\end{itemize}
Therefore, each of the nine regions in the quench parameters space is associated, in the GGE, to an effective metallic or insulating ``phase", and transitions occur whenever one of the $\delta_i$ crosses the critical lines.
\subsection{A geometrical interpretation}\label{SM:subsec:geom}
\begin{figure}[h]
	\centering
	\includegraphics[width=1\linewidth]{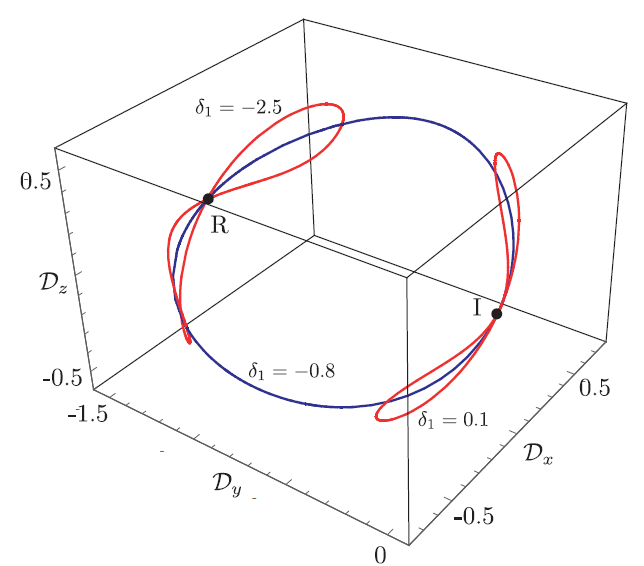}
	\caption{Three different curves $ \gamma $ described by the vector $ \vec{\mathcal{D}}_k $ with $ k $ spanning the BZ. Here we set $ \delta_0=5w $ and three possible representations of the different regimes are shown for $ \delta_1=0.1w $ and $ \delta_1=-2.5w $, which describe the insulating effective phase (red curves), and for $ \delta_1=-0.8w $, which describe the metallic effective phase (blu curve).}
	\label{SM:fig:geometry}
\end{figure}

We can provide a geometrical interpretation of what discussed above. As $k$ sweeps the BZ, the vector $\vec{\mathcal D}_k$ describes a closed curve $\gamma$ in the three-dimensional space, pinned to either or both the fixed points I, R. To be specific and without loss of generality, here we consider the case $\delta_0>0$ only. 
\begin{itemize}
	\item For $\delta_1>0$, one has $\Delta_0=\Delta_\pi=-1$. Thus, $\gamma$ passes twice through the point I;	
	\item For $-2<\delta_1<0$, one has $\Delta_0=-\Delta_\pi=-1$. As a consequence, $\gamma$ passes through both I and R;	
	\item For $\delta_1<-2$ and $\gamma$ passes through both I and R;	
\end{itemize}
In the first and last cases the GGE has an insulating character and the curve $\gamma$ describes a butterfly shape pinned either at the origin ($\delta_1>0$) or at $(0,-\pi/2,0)$ ($\delta_1<-2$). On the other hand, for $-2<\delta_1<0$ the GGE is metallic and $\gamma$ describes a closed loop pinned at I and R. A variation of the quench parameters which does not result in a crossing of the critical lines does not alter the qualitative features of the curve $\gamma$. The scenario is summarized in Fig.~\ref{SM:fig:geometry}.
\subsection{Robustness}\label{SM:subsec:Stab}
In this section we give some details about the robustness of the effective metal-insulator transition with respect to a thermal initial state and a finite-duration quench.\\
\subsubsection{Thermal initial state}
We start by observing that, given a generic occupation of the pre-quench states $n_{\alpha,k}^{(0)}$, using the quench transformation one can promptly obtain
\begin{equation}
n_{c,k}-n_{v,k}=[n_{v,k}^{(0)}-n_{c,k}^{(0)}]\Delta_k\,.	
\end{equation}
This relationship is very powerful: Let us apply it to the case of a system prepared at a generic temperature $ T=(k_B\ \beta)^{-1}$, with occupation numbers
\begin{equation}
n_{c,k}^{(0)}=\dfrac{1}{1+e^{\beta\epsilon_{0,k}}} \qquad \text{and} \qquad n_{v,k}^{(0)}=\dfrac{1}{1+e^{-\beta\epsilon_{0,k}}}\,.
\end{equation}
One then easily obtains
\begin{equation}
\label{eq:DeltaT}
n_{c,k}-n_{v,k}=f_{T,k}\Delta_k\,,
\end{equation}
with
\begin{equation}\label{SM:eq:thermal}
f_{T,k}=\dfrac{\sinh(\beta\epsilon_{0,k})}{1+\cosh(\beta\epsilon_{0,k})}>0 .
\end{equation}
Equipped with Eq.~(\ref{eq:DeltaT}) one obtains the GGE multipliers 
\begin{equation}
\lambda_{c,k}= \ln\left(\dfrac{1-f_{T,k}\Delta_k}{1+f_{T,k}\Delta_k}\right) \qquad \text{and} \qquad
\lambda_{v,k}=-\lambda_{c,k}\,.
\end{equation}
The key observation is that for any temperature $T$ the qualitative features of $n_{c,k}-n_{v,k}$ (governing the inversion of population) and of the new GGE multipliers (dictating the effective metal-insulator transitions) remain unchanged since $f_{T,k}$ has no zeroes and thus can neither destroy the insulating phase, nor distort the metallic one. Thus, all the conclusions obtained in the $ T=0 $ case still hold, including the presence of non-analyticities in the steady state of quantities, still hold.
\subsubsection{Finite-duration quenches}
Turning to the study of the effects of a quench with finite duration, we introduce a new quench protocol encoded in the time dependence of the quench parameter $ \delta(t) $, defined in Eq.~\eqref{SM:eq:H(t)}.
Here we consider a linear ramp, namely 
\begin{equation}
\delta(t)=\left\{\begin{array}{l}
\delta_0  \\ 
\delta_0+(\delta_1-\delta_0)t/\tau  \\ 
\delta_1 
\end{array} \right. \begin{array}{l}
t\leq 0\\ 
0<t\leq\tau \\ 
t>\tau
\end{array} ,
\end{equation}
where $ \tau $ is the quench duration. By means of the Heisenberg equation of motion and taking the following ansatz~\cite{noi,gge9}, 
\begin{align}
\begin{split}
\left(\begin{array}{c}
c_{k,A}(t)\\ 
c_{k,B}(t)
\end{array} \right)=&\left(\begin{array}{cc}
f_{k,A}(t) & g_{k,A}(t) \\ 
f_{k,B}(t) & g_{k,B}(t)
\end{array} \right) \left(\begin{array}{c}
c_{k,A}\\ 
c_{k,B}
\end{array} \right)  \\ =&\ V_k(t)\left(\begin{array}{c}
c_{k,A}\\ 
c_{k,B}
\end{array} \right) ,
\end{split}
\end{align}
where $ c_{k,A} $ and $ c_{k,B} $ are the Fermi operator in the Schr\"odinger picture at $ t=0 $,
we can evaluate the time evolution of the whole Fermi spinor $  \Psi_k^\dagger= \left(c_{k,A}^\dag,c_{k,B}^\dag \right) $, given the initial conditions $ f_{k,A}(t=0)=g_{k,B}(t=0)=1 $ and $ f_{k,B}(t=0)=g_{k,A}(t=0)=0 $. We obtain that the coefficients of the matrix $ V_k(t) $ satisfy the following systems of differential equations:
\begin{align}
\begin{split}
\left(\begin{array}{c}
f_{k,A}(t)\\ 
f_{k,B}(t)
\end{array} \right)=\left(\begin{array}{cc}
0 & m_{k}(t) \\ 
m_{k}^*(t) & 0
\end{array} \right) \left(\begin{array}{c}
f_{k,A}\\ 
f_{k,B}
\end{array} \right) , \\ \left(\begin{array}{c}
g_{k,A}(t)\\ 
g_{k,B}(t)
\end{array} \right)=\left(\begin{array}{cc}
0 & m_{k}(t) \\ 
m_{k}^*(t) & 0
\end{array} \right) \left(\begin{array}{c}
g_{k,A}\\ 
g_{k,B}
\end{array} \right) ,
\end{split}
\end{align}
where \begin{equation}
m_k(t)=1+e^{-i k}+\delta(t) .
\end{equation}
Therefore, we are able to write the transformation which connects the pre- and post-quench diagonal bases, 
\begin{equation}\label{SM:eq:Mtrasf}
\Phi_{1,k}= U_{1,k} V_k(\tau) U_{0,k}^\dag \Phi_{0,k} =\mathcal{V}_{1,k}^0(\tau) \Phi_{0,k} .
\end{equation} 
which represents the generalization to the finite duration quench of Eq.~\eqref{SM:eq:Transfbases}. Equation~\eqref{SM:eq:Mtrasf} allows us to evaluate the GGE conserved quantities
\begin{align}
\begin{split}
n_{c,k}=&\left|A_{0,k}\left[B_{1,k}g_{k,B}(\tau)+A_{1,k}g_{k,A}(\tau)\right]\right.\\&\left.-B_{0,k}\left[B_{1,k}f_{k,B}(\tau)+A_{1,k}f_{k,A}(\tau)\right]\right|^2 = 1-n_{v,k}.
\end{split}
\end{align}
As done above, we focus now on the points $k=0,\pi$, where the analysis becomes transparent. At these points $m_{k}(t)$ is real and the coefficients of the matrix $ V_k(t) $ fulfill the following differential equation
\begin{equation}
\partial_t^2 V_k-\dfrac{\delta_1-\delta_0}{\tau \mu_k(t)} \partial_t V_k+\mu_k(t)^2 V_k =0 ,
\end{equation} 
where $ \mu_0(t)=2+\delta(t) $ and $ \mu_\pi(t)=\delta(t) $. Solving this equation and assuming for simplicity $ \delta_0>0 $, we obtain for $ k=\pi $
\begin{equation}
\mathcal{V}_{1,\pi}^0(\tau)=\dfrac{1+s(\delta_1)}{2} \left(\begin{array}{cc}
e^{-i \eta \tau} & 0 \\ 
0 & e^{i \eta \tau}
\end{array} \right) -\dfrac{1-s(\delta_1)}{2} \left(\begin{array}{cc}
0 &e^{i \eta \tau} \\ 
- e^{-i \eta \tau} & 0
\end{array} \right) ,
\end{equation}
where $ \eta=\dfrac{\delta_0+\delta_1}{2} $. Analogous results are achieved for $ k=0 $, where the sign function is shifted to the second critical point, i.e. it becomes $ \text{s}(\delta_1+2)$. In qualitative agreement with the sudden case, to which the above equation reduces for $\eta\to 0$, for $\delta_1>0$ the $c,v$ bands remain essentially the same with the exception of an $\eta$-dependent phase shift while for $\delta_1<0$, in addition to the $c,v$  the $\eta$-dependent phase shift, the $c,v$ bands swap their role. Crucially, however, the phase shift is irrelevant in the evaluation of $\Delta_k$ at $k=0,\pi$. As a consequence, the same qualitative conclusions concerning a non-trivial inversion of population and an effective metal-insulator transition can be drawn.
\subsection{Other observables}\label{SM:subsec:Obs}
In this section we provide some details about the steady state value of the quantities discussed in the main text.
\subsubsection{Dimerization}
The dimerization operator is defined as \cite{Asboth}
\begin{equation}
\mathcal{M}(x)=\sum_{k,k'}e^{i(k'-k)x}\Psi^{\dagger}_{k}\sigma_x\Psi_{k'} .
\end{equation}
Exploiting Eqns.~(\ref{eq:S5}-\ref{eq:S7}) and the definition of the GGE one finally obtains
\begin{align}
\begin{split}
\bar{\mathcal{M}}&= \braket{G_0|M(x)|G_0} \\ &=\dfrac{1}{\pi} \int_{-\pi}^\pi  A_{1,k}\text{Re}\{B_{1,k}\} \left(n_{c,k}-n_{v,k}\right) dk .
\end{split}
\end{align}
\subsubsection{Entropy}\label{SM:subsec:Entropy}
We consider the entropy associated to the GGE, defined as
\begin{equation}
S=\tr\left\{\rho_{G}\ln(\rho_{G})\right\} .
\end{equation}
The quantity $ S $ has to be interpreted as the extensive part of the entanglement
entropy of long enough subsystems. Its evaluation can be performed by
standard means by noticing the formal analogy to a system of free fermions unveiled in Sec.~\ref{SM:subsec:GGE}. We obtain:
\begin{equation}
S=-\sum_{\alpha,k} n_{\alpha,k}\ln\left(n_{\alpha,k}\right)\,.
\end{equation}
\subsubsection{Current fluctuations}\label{SM:subsec:CurrFluc}
Here we consider the fluctuations of the spatially-averaged current, related to the DC conductance, both in the real and in the effective GGE bands. We start by defining the current operator as the derivative of the energy spectrum with respect to the momentum $ k $. In the thermodynamic limit, 
\begin{equation}
J_{(0)}=\sum_{\alpha} J_\alpha^{(0)}=\dfrac{1}{2\pi}\sum_{\alpha}\int_{-\pi}^{\pi} j_{\alpha,k}^{(0)} N_{\alpha,k} dk
\end{equation}
with $ j_{\alpha,k}=\pm\partial_k \epsilon_{1,k} $ ($\pm$ for $c,v$) and $ j_{\alpha,k}^0= \partial\varepsilon_{\alpha,k}$ for the real $ (J) $ and GGE effective bands $ (J_0) $ respectively. The DC fluctuations $\bar\sigma_{(0)}=\langle G_0|J_{(0)}^2|G_0\rangle$ are thus given by
\begin{equation}
\bar\sigma_{(0)}=\dfrac{1}{(2\pi)^2} \sum_{\alpha} \int_{-\pi}^{\pi} \left(j^{(0)}_{\alpha,k}\right)^2 n_{\alpha,k}\left(1-n_{\alpha,k}\right)dk
\end{equation}
and, given the relation between the GGE conserved quantities and the effective bands in Eq.~\eqref{SM:eq:Delta}, we obtain
\begin{equation}
\bar\sigma_{(0)}=\dfrac{1}{(2\pi)^2} \int_{0}^{\pi} \left(j^{(0)}_{c,k}\right)^2 \dfrac{dk}{1+\cosh(\lambda_{c,k})} .
\end{equation}
\section{Interacting model}\label{SM:sec:Interactions}
\begin{figure}[h]
	\centering
	\includegraphics[width=1\linewidth]{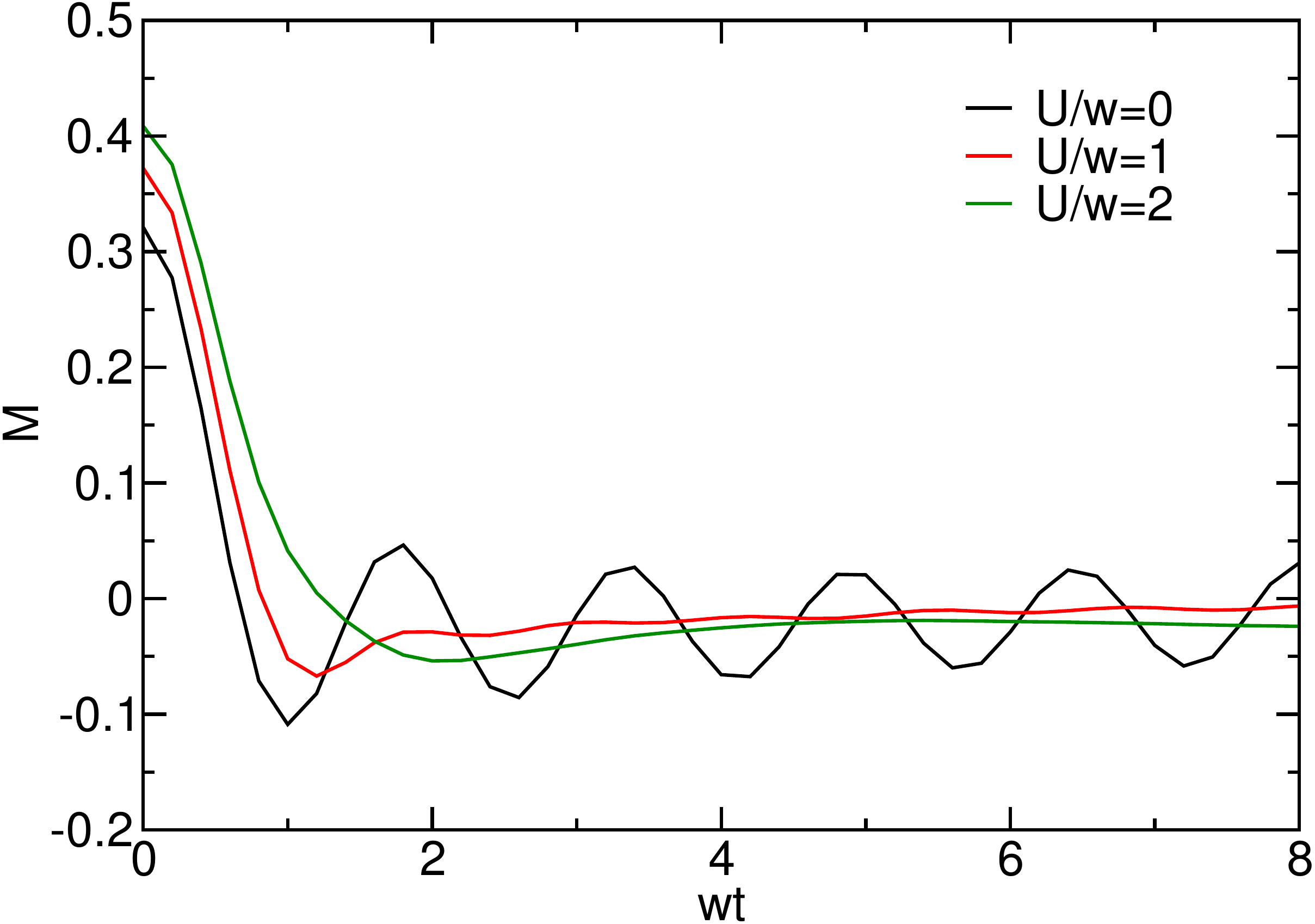}
	\caption{Time evolution of $ M $ for different values of the interaction strength $ U $, for $ \delta_0=w $ and $ \delta_1=-0.01w $. Increasing the interaction strength strongly suppresses the 
		transient oscillations. }
	\label{SM:fig:Quench_dynamics}
\end{figure}

To include interactions we concentrate on a model that is easier to simulate than the SSH model considered in the rest of the text. The Hamiltonian we want to consider is given by spinless fermions on a chain where we allow for a staggered field, nearest neighbor hopping as well as interactions 
\begin{equation}
H(t)=\sum_{i=1}^{N} w c_i^\dagger c_{i+1}+{\rm H.c.}+\delta(t)(-1)^i n_i+Un_{i}n_{i+1}.
\end{equation} 
Here $c_i^{(\dagger)}$ annihilates (creates) a spinless fermion on lattice site $i$. The quench is performed at time $t=0$, at which the staggered field is is abruptly changed from $\delta_0$ to $\delta_1$. This model exhibits the same qualitative behavior of the SSH model in the absence of interactions. To simulate the dynamics we use an implementation of the density matrix renormalization group directly set up in the thermodynamic limit $N\to\infty$. Here we use an iterative algorithm to prepare the ground state of $H(t<0)$ and then propagate the wave function in real time with respect to $H(t>0)$ employing a fourth order Suzuki-Trotter decomposition. The decomposition time steps are chosen small enough to yield converged results.  We dynamically increase the so-called bond-dimension, the parameter describing the numerical accuracy, as the simulation time progress, which allows us to achieve numerically exact results. By this procedure the truncation error in the wavefunction is kept below a $10^{-7}$ threshold.  As simulation time progresses the entanglement in the system rises and with it the bond dimension as well as the numerical effort needed. Entanglement growth in simulation time is typically linear leading to an exponential increase in bond dimension.  Therefore, at a certain time $t$ the numerical resources are exhausted and no further progress in simulation time can be made. Luckily, at finite $U$ the dynamics for the observable $\bar{\mathcal{M}}=\left\langle n_0-1/2\right\rangle$ of interest become strongly damped facilitating  an extrapolation to long times, compare Fig.~\ref{SM:fig:Quench_dynamics}.  For $U=0$, where the strong oscillations make an extrapolation more difficult, we check convergence by comparison with exact results obtained from the GGE directly.

\begin{thebibliography}{99}
\bibitem{FN0} This inversion of population also gives rise to a dynamical quantum phase transition~\cite{heyl}, which we observe in the time evolution of the Loschmidt echo of the system. This feature, however, does not seem to show up into the dynamics of observable quantities.
\bibitem{FN2} Here, the current density operator has been written for $x=0$ since we are only interested in fluctuations at equal space.
\bibitem{FN3} The initial state is prepared at a finite temperature $T$, but no contact with an external thermostat is assumed during or after the quench.
\bibitem{FN4} Similarly, other reports of the intrinsic stability of quantum quench effects with respect to a thermal preparation have been published~\cite{scipost}.
\end{thebibliography}
\end{document}